\documentclass[aps,pre,showpacs,amsmath,amssymb,longbibliography]{revtex4-2}
\usepackage{xcolor,hyperref}
\hypersetup{
	colorlinks,
	linkcolor={blue!50!black},
	citecolor={blue!50!black},
	urlcolor={blue!80!black}
}
\usepackage{graphicx,float}
\usepackage{xcolor}
\newcommand{\bF}{\bold F}

\newcommand{\bR}{\bold R}
\newcommand{\br}{\bold r}

\newcommand{\be}{\begin{equation}}
	\newcommand{\ee}{\end{equation}}
\newcommand{\fig}[1]{Fig.~\ref{#1}}
\newcommand{\Fig}[1]{Figure~\ref{#1}}

\newcommand{\eq}[1]{Eq.~(\ref{#1})}

\newcommand{\bo}{\mathbf n}
\newcommand{\bxi}{\boldsymbol{\xi}}
\graphicspath{ {./figs/} }

\begin{document}
	\title{Active-parameter polydispersity in the 2d ABP Yukawa model}
	\date{\today}
	\author{Shibu Saw}\email{shibus@ruc.dk}
	\affiliation{\textit{Glass and Time}, IMFUFA, Department of Science and Environment, Roskilde University, P.O. Box 260, DK-4000 Roskilde, Denmark}
	\author{Lorenzo Costigliola}\email{lorenzoc@ruc.dk}
	\affiliation{\textit{Glass and Time}, IMFUFA, Department of Science and Environment, Roskilde University, P.O. Box 260, DK-4000 Roskilde, Denmark}
	\author{Jeppe C. Dyre}\email{dyre@ruc.dk}
	\affiliation{\textit{Glass and Time}, IMFUFA, Department of Science and Environment, Roskilde University, P.O. Box 260, DK-4000 Roskilde, Denmark}

	\begin{abstract}
		In both experiments and simulations the most commonly studied kind of parameter polydispersity is that of varying particles size. This paper investigates by simulations the effects of introducing polydispersity in other parameters for two-dimensional Active Brownian Particles with Yukawa pair interactions. Polydispersity is studied separately in the translational and rotational diffusion coefficients, as well as in the swim velocity $v_0$. Uniform and binary parameter distributions are considered in both the homogeneous and the motility-induced phase-separation (MIPS) phases. We find only minute changes in structure and dynamics upon the introduction of parameter polydispersity, even for situations involving 50\% polydispersity. The reason for this is not clear. An exception is the case of $v_0$ polydispersity for which the average radial distribution function with changing polydispersity shows significant variations in the MIPS phase. Even in this case, however, the dynamics is only modestly affected. As a possible application of our findings, we suggest that a temporary introduction of polydispersity into a single-component active-matter model characterized by a very long equilibration time, i.e., a glass-forming active system, may be used to equilibrate the system efficiently by particle swaps.
	\end{abstract}
	
	\maketitle

	\section{Introduction}

	Active matter includes fluids of self-propelled particles like bacteria, birds, or insect flocks  \cite{ang11a,mar13a,bec16,ram17,sai18,sha21,bow22}. An example of the intriguing features of active matter is motility-induced phase separation (MIPS), the fact that a purely repulsive system may phase separate into high- and low-density phases \cite{das14,cat15,ram17,gey19,das20,mer20}. 
	
	There is currently a considerable interest in passive polydisperse systems, in particular deriving from the use of polydispersity for SWAP-equilibrating models of supercooled liquids \cite{nin17}. An obvious question that arises is: how different are the dynamics of the different particles \cite{abr08b,zac15,pih23}? Polydispersity is also relevant for active matter models because for a biological system one cannot expect all constituents to be identical \cite{ni15,hen20,kum21,sza21}. Models with motility polydispersity are relevant for both biological and colloidal active systems; thus Castro {\it et al.} recently showed that the MIPS phase gets suppressed with the introduction of a spread of swim (self-propelled) velocities in the Active Brownian Particles (ABP) model \cite{solich21}. 
	
	This paper presents a systematic study of the effects of polydispersity in other parameters than size \cite{ket22,deb23} of the ABP model in two dimensions. The particles interact via the Yukawa (screened Coulomb) pair potential \cite{yuk35,han13}, and polydispersity is introduced by varying the three activity parameters controlling the motion of the individual particles. We find a surprisingly small effect of even quite high polydispersity, up to 50\%, when parameters vary such that their average is kept constant. This applies to both continuous and binary polydispersity and is in sharp contrast to the large effects of size polydispersity \cite{ket22,deb23}.

	\section{The 2d ABP Yukawa system}
	
	The Yukawa pair potential \cite{yuk35,mea21} is defined \cite{saw23b} by
	
	\be\label{yuk}
	v(r)
	\,=\,\frac{Q^2\sigma}{r}\,\exp\left(-\frac{r}{\lambda\sigma}\right)\,.
	\ee
	Here $\sigma$ is a length parameter, $\lambda$ is dimensionless, and the ``charge'' $Q$ has dimension square root of energy. Throughout the paper we use the fixed values $\lambda=0.16$ and $Q=50$, while $\sigma\equiv 1$ defines the unit of length and thus the unit of particle density. 
	
	If $\br_i$ is the position vector of particle $i$, the ABP equations of motion in two dimensions are \cite{far15}
	
	\be\label{ABP_EOM2}
	\dot{\br}_i
	\,=\,\mu\bF_i+\bxi_i(t)+v_0\, \bo_i(t)\,.
	\ee
	Here, $\mu$ is the mobility (velocity over force), $\bF_i(\bR)=-\nabla_i U(\bR)$ is the force on particle $i$ in which $\bR=(\br_1,....,\br_N)$ is the configuration vector and $U(\bR)=\sum_{i<j}v(r_{ij})$ (sum over all particle pairs) is the potential-energy function, $\bxi_i(t)$ is a Gaussian random white-noise vector, and $v_0$ is the ``swim velocity''. The vector $\bo_i(t)=(\cos\theta_i(t),\sin\theta_i(t))$ is a stochastic unit vector in which the angle $\theta_i(t)$ is controlled by a Gaussian white noise term the magnitude of which defines the rotational diffusion coefficient, $D_r$, according to
	
	\be\label{oi_noise}
	\langle\dot\theta_i(t)\dot\theta_j(t')\rangle
	\,=\,2D_r\delta_{ij}\,\delta(t-t')\,.
	\ee
	The magnitude of the white-noise velocity vector $\bxi_i(t)$ defines the translational diffusion coefficient $D_t$,
	
	\be\label{xi_noise2}
	\langle\bxi_i^\alpha(t)\bxi_j^\beta(t')\rangle
	\,=\,2D_t\delta_{ij}\delta_{\alpha\beta}\delta(t-t')
	\ee
	in which $\alpha,\beta$ are spatial $x,y$ indices and $i,j$ are again particle indices. The mobility $\mu$ is taken to be unity throughout, i.e., $\mu$ is regarded as a material constant, while the remaining model parameters $D_r$, $D_t$, and $v_0$ are allowed to vary from particle to particle. This introduces three kinds of polydispersity. In all cases the average of the polydisperse parameter is kept constant. For any varying parameter $X$, the polydispersity $\delta$ is conventionally defined \cite{fre86} as $\delta\equiv\sqrt{\langle X^2\rangle -\langle X\rangle^2}/\langle X\rangle$ in which the sharp brackets denote averages.
	
	We simulated $10000$ particles of the 2d Yukawa system with interactions cut off at $4.5\sigma$. The time step used was $\Delta t=0.0625 \langle D_t\rangle/\langle v_0\rangle^2$. Each simulation involved $2 \cdot 10^7$ time steps. The (GPU) code employed was RUMD \cite{RUMD}, modified to deal with polydispersity in particle-activity parameters. Parameters corresponding to both the homogeneous phase ($D_t=1.0$, $D_r=0.8$, $v_0=25$) and the MIPS phase ($D_t=1.0$, $D_r=0.2$, $v_0=25$) were simulated.

	\section{Parameter polydispersity in the homogeneous phase}
	
	\begin{figure}[H]
		\begin{center}
			\includegraphics[width=0.98\textwidth, angle=0]{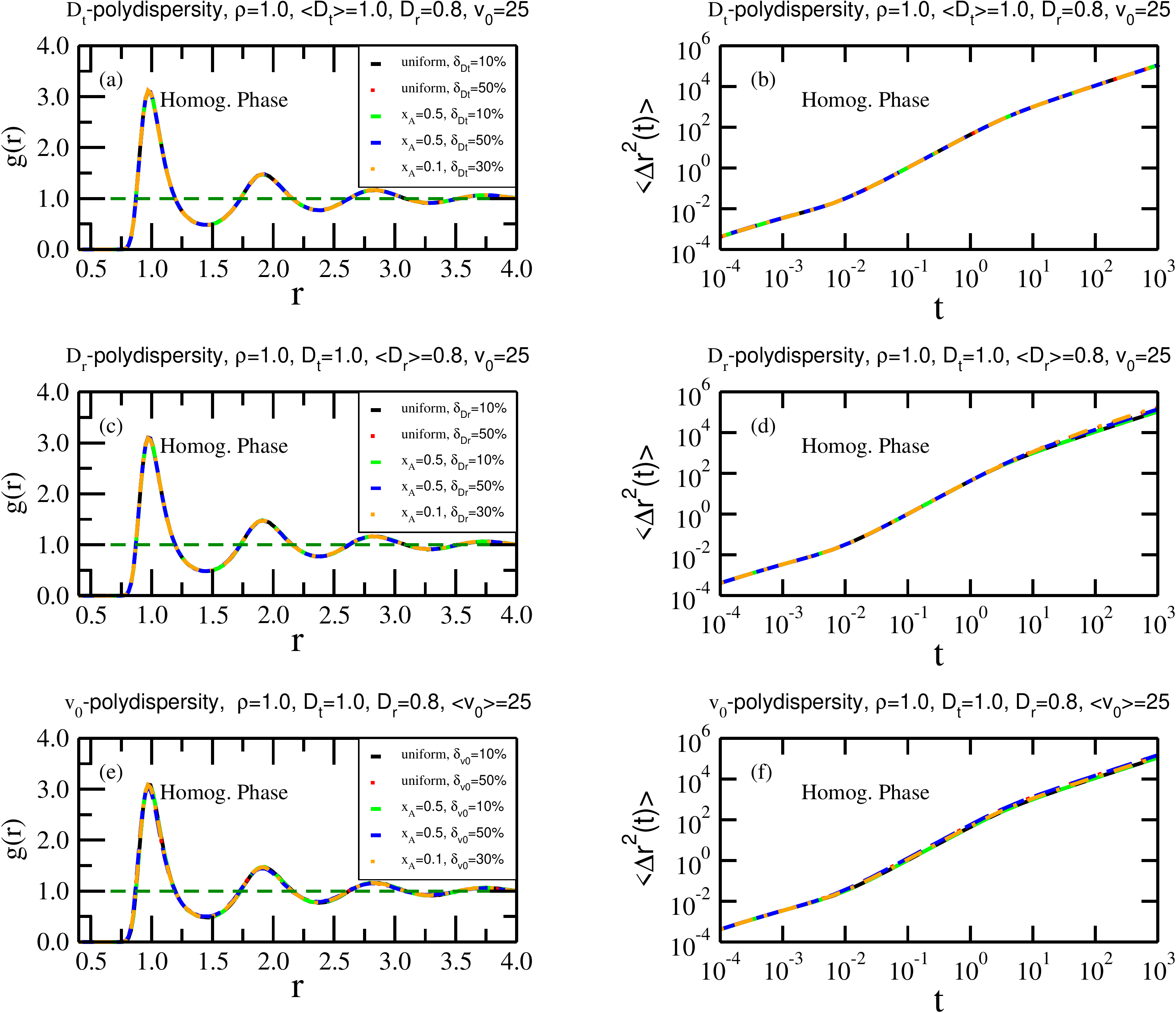}\vspace{2mm}
			\caption{Structure and dynamics in the homogeneous phase for uniformly polydisperse systems (black and red curves are for 10\% and 50\% polydispersity) and for binary systems in which $x_A$ is the fraction of large-parameter particles (green represents 10\%, orange 30\%, and blue 50\% polydispersity). 
				(a), (c), and (e) show the average radial distribution functions (RDFs), $g(r)$, for systems with polydispersity in the $D_t$, $D_r$, and $v_0$ parameters, respectively. 
				(b), (d), and (f) show the corresponding results for the mean-square displacement, MSD, as a function of time, $\langle\Delta r^2(t)\rangle$. 
				In all cases there is little effect of polydispersity.}
			\label{fig1}
		\end{center}
	\end{figure}

	We first consider the effect of active-parameter polydispersity on the structure and dynamics in the homogeneous phase. \Fig{fig1}(a), (c), and (e) show the (average) radial distribution functions (RDFs) for different degrees of polydispersity: uniform parameter distributions of 10\% and 50\% polydispersity and binary parameter distributions of 10\%, 30\%, and 50\% polydispersity in $D_t$ ((a) and (b)), $D_r$ ((c) and (d)), and $v_0$ ((e) and (f)) ($x_A$ denotes the large-parameter fraction of particles). 
	
	\Fig{fig1}(b), (d), and (f) show the average mean-square displacement (MSD) as a function of time for the same situations. We find here the well-known three regimes \cite{cap21a}: diffusive (small time), ballistic (intermediate time), and diffusive (long time). The first regime is governed by the thermal noise, the second by the swim velocity, and the third by the rotational diffusion coefficient and swim velocity. In all cases there is little effect of polydispersity. This is not trivial because the individual particles conform to different equations of motion; indeed they move differently as becomes clear from the next figure.

	\begin{figure}[H]
		\begin{center}
			\includegraphics[width=0.98\textwidth, angle=0]{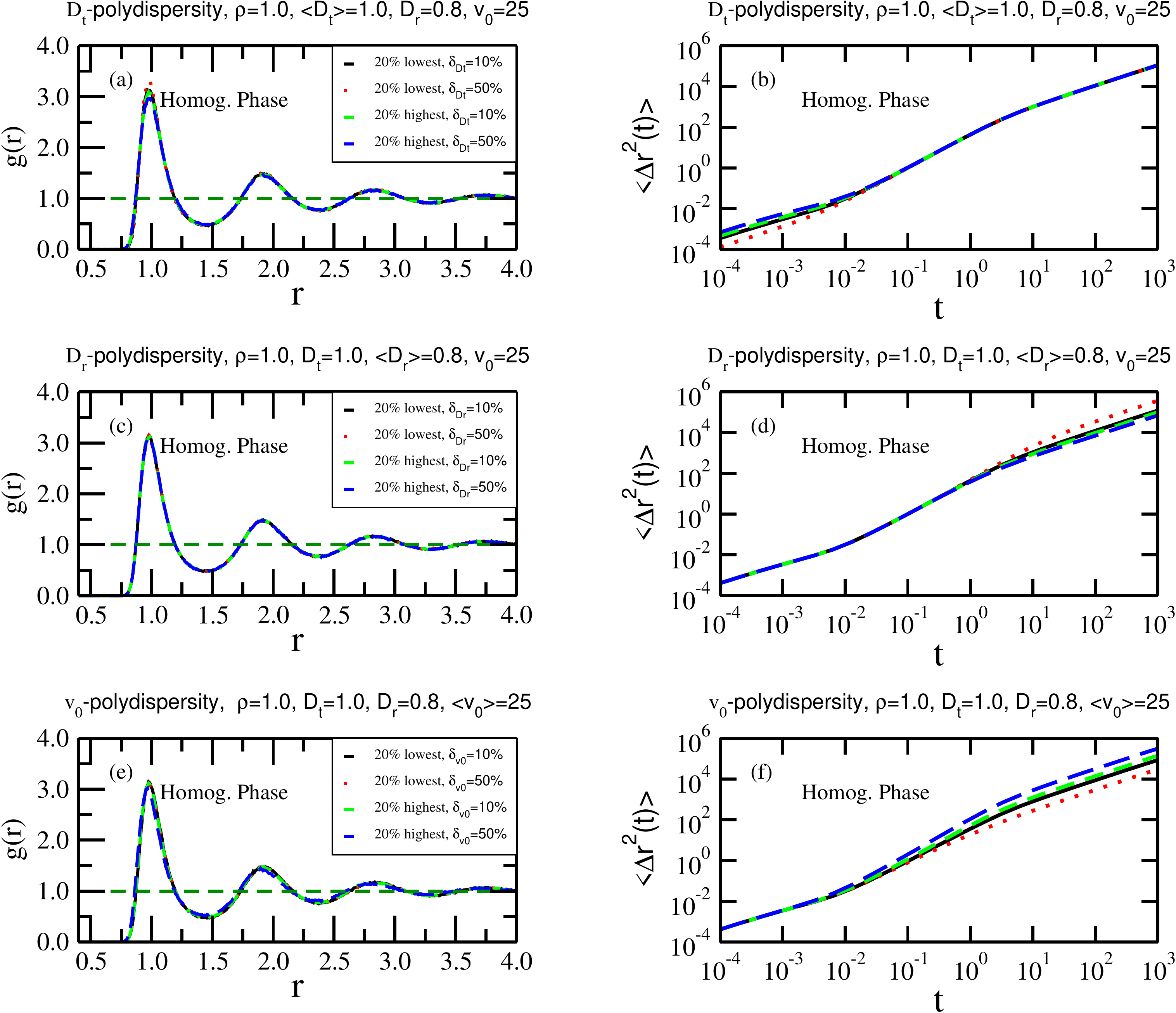}\vspace{2mm}
			\caption{Role of the 20\% lowest (black and red) and 20\% highest (green and blue) active-parameter particles in the homogeneous phase of continuously polydisperse systems at 10\% and 50\% polydispersity. 
				(a), (c), and (e) show RDFs, $g(r)$, for polydispersity in the $D_t$, $D_r$, and $v_0$ parameter, respectively. 
				(b), (d), and (f) show the corresponding results for the MSD $\langle\Delta r^2(t)\rangle$. 
				For the RDFs there is litte difference between the lowest and highest active-parameter particles, while the MSD shows variations that are much larger than those of the overall average (\fig{fig1}). This variation is seen in the short-time data in the case of $D_t$ polydispersity and in the long-time data for $D_r$ and $v_0$ polydispersity.}
			\label{fig2}
		\end{center}
	\end{figure}

	To illuminate the role of parameter polydispersity for the individual particles, we identified for the two uniform polydispersities the particles with the 20\% lowest activity parameters and those with the 20\% highest. For each of these categories we determined the corresponding RDFs (counting only surrounding particles of the same type) and MSDs. The results are shown in \fig{fig2}. For the structure ((a), (c), and (e)) there is little difference although one notes that the low $D_t$ particles show a somewhat more pronounced first peak than that of the high $D_t$ particles ((a)). Since $D_t$ in the Langevin-type equation \eq{ABP_EOM2} plays the role of a temperature, this is consistent with the well-known finding for passive systems that lowering the temperature generally leads to a higher first peak of the RDF. For the dynamics, there are clear differences: In the case of $D_t$ polydispersity ((b)), the long-time dynamics is the same, while the short-time dynamics is fastest for the highest $D_t$ particles. For $D_r$ polydispersity the opposite is observed; here the short-time dynamics is the same for low and high $D_r$ particles while the long-time dynamics is fastest for the particles with low $D_r$. The rotational diffusion coefficient determines a particle's persistence time because a decrease of $D_r$ implies an increase of the long-time diffusion coefficient. Thus the phenomena observed are consistent with the single-particle scaling of the long-time diffusion coefficient. On the short time scale little change of direction is possible, making the value of $D_r$ irrelevant. 
	
	Consider finally the case of $v_0$ polydispersity ((f)). Here there is no effect on the short-time dynamics, while the long-time dynamics is fastest for particles with large $v_0$. That the short-time dynamics is unaffected is a simple consequence of \eq{ABP_EOM2} in which the $v_0$ term on the short time scale gives rise to a MSD proportional to $t^2$, which is much smaller than the short-time diffusive contribution to the MSD. The faster long-time diffusive dynamics for the high $v_0$ particles comes about because a higher swim velocity implies larger displacements in one direction before the direction changes, corresponding to longer jumps in a simple random-walk picture.
	
	How do the results of \fig{fig2} relate to the overall average structure and dynamics findings of \fig{fig1}? The structure is almost the same for low and high active parameter particles for all three types of polydispersity -- and independent of the degree of polydispersity -- so in this light the RDF findings of \fig{fig1} are not surprising. In regard to the MSD, however, the variations induced by parameter polydispersity are significant but strikingly average out, resulting in little overall change of the average MSD. Thus in all three cases the black and green curves in \fig{fig2}, which represent just 10\% polydispersity, are close to each other, while the red and blue curves (50\% polydispersity) move in opposite directions.

	\section{Parameter polydispersity in the MIPS phase}
	
	The existence of a MIPS phase is a unique feature of active matter, and MIPS is also found in the ABP model \cite{dig18,cap20b,cap20d}. This phase is also of interest to investigate in regard to the effects of introducing active-parameter polydispersity. We did this by repeating the simulations, the only difference being that the average of $D_r$ is now 0.2 instead of the above used 0.8. The majority of particles are found in the dense phase at all MIPS state points studied, meaning that data found by averaging over all particles are representative for this phase.

	\begin{figure}[H]
		\begin{center}
			\includegraphics[width=0.98\textwidth, angle=0]{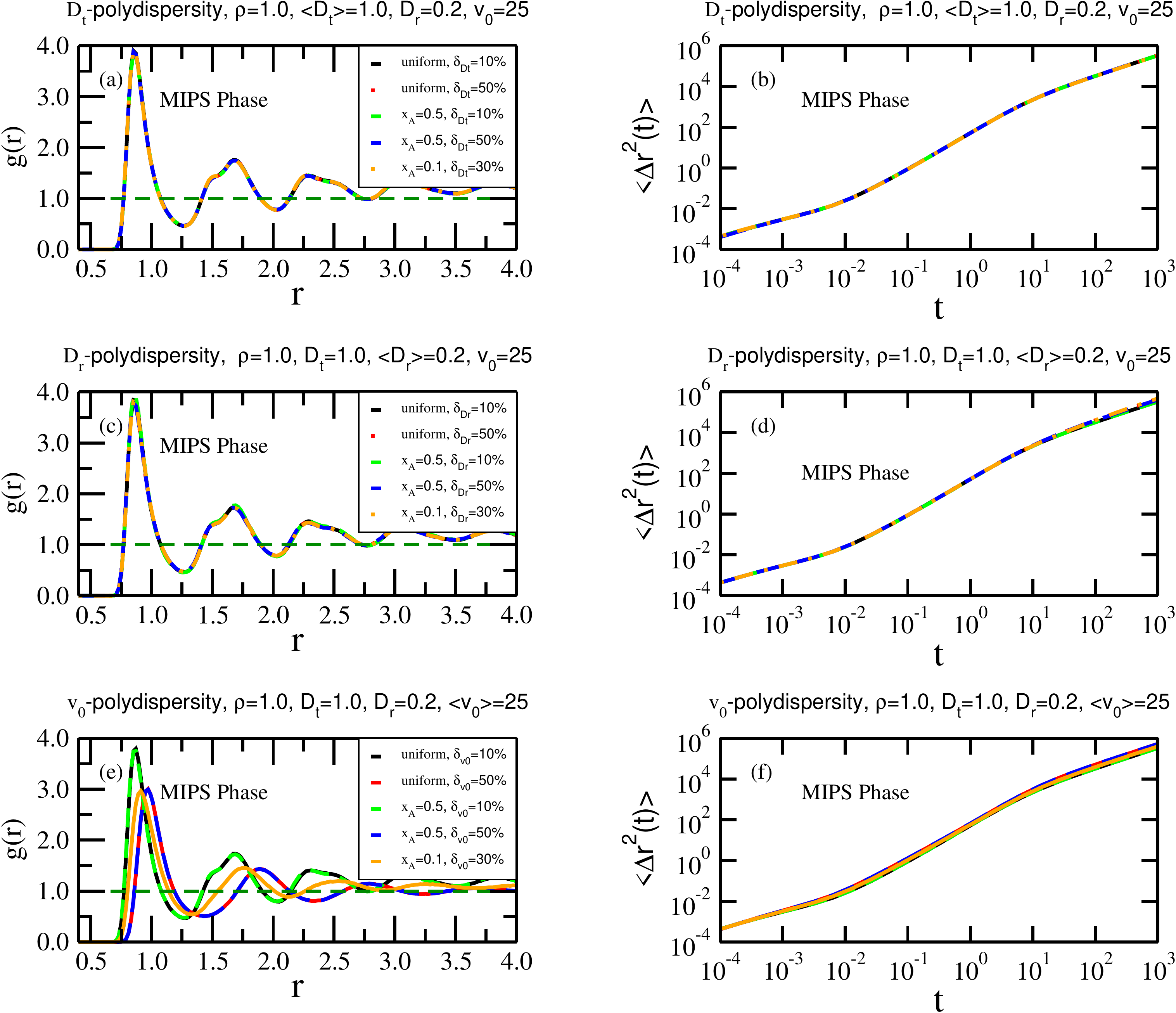}\vspace{2mm}
			\caption{Structure and dynamics in the MIPS for uniformly polydisperse systems (black and red curves are for 10\% and 50\% polydispersity) and for binary systems in which $x_A$ is the fraction of large-parameter particles (green represents 10\%, orange 30\%, and blue 50\% polydispersity). 
				(a), (c), and (e) show the average RDFs for systems with polydispersity in the $D_t$, $D_r$, and $v_0$ parameters, respectively. 
				(b), (d), and (f) show the corresponding results for the mean-square displacement as a function of time. There is little effect of introducing polydispersity in the $D_t$ and $D_r$ parameters whereas a notable effect of $v_0$ polydispersity is observed for the RDF, in which case there is also a visible -- though much smaller -- effect on the dynamics.}
			\label{fig3}
		\end{center}
	\end{figure}

	The results for the RDFs and MSDs are shown in \fig{fig3}. In regard to the dynamics, the picture is not much different from that of the homogeneous phase: the MSD is virtually unaffected by the introduction of polydispersity in the three parameters ((b), (d), and (f)). The same applies for the RDF for $D_t$ and $D_r$ polydispersity, whereas $v_0$ polydispersity strongly affects the RDF ((e)). Note that the RDF at large $r$ is systematically slightly larger than unity; this is an effect of the fact that the figures report the RDF averaged over all particles. While not clearly visible, a close inspection reveals that the green RDF and MSD curves cover a black one and the blue curves likewise cover a red one. The former are for 10\% polydispersity in the uniform and binary cases, respectively, while the latter are for 50\% polydispersity. We conclude that the introduction of $v_0$ polydispersity strongly affects the RDF in a way that is independent of the parameter probability distribution. Given that the existence of the MIPS phase reflects the active-matter feature of a temporary persistence direction in the particle motion, it is not surprising that introducing $v_0$ polydispersity has a strong effect on the structure of the MIPS phase.

	\begin{figure}[H]
		\begin{center}
			\includegraphics[width=0.98\textwidth, angle=0]{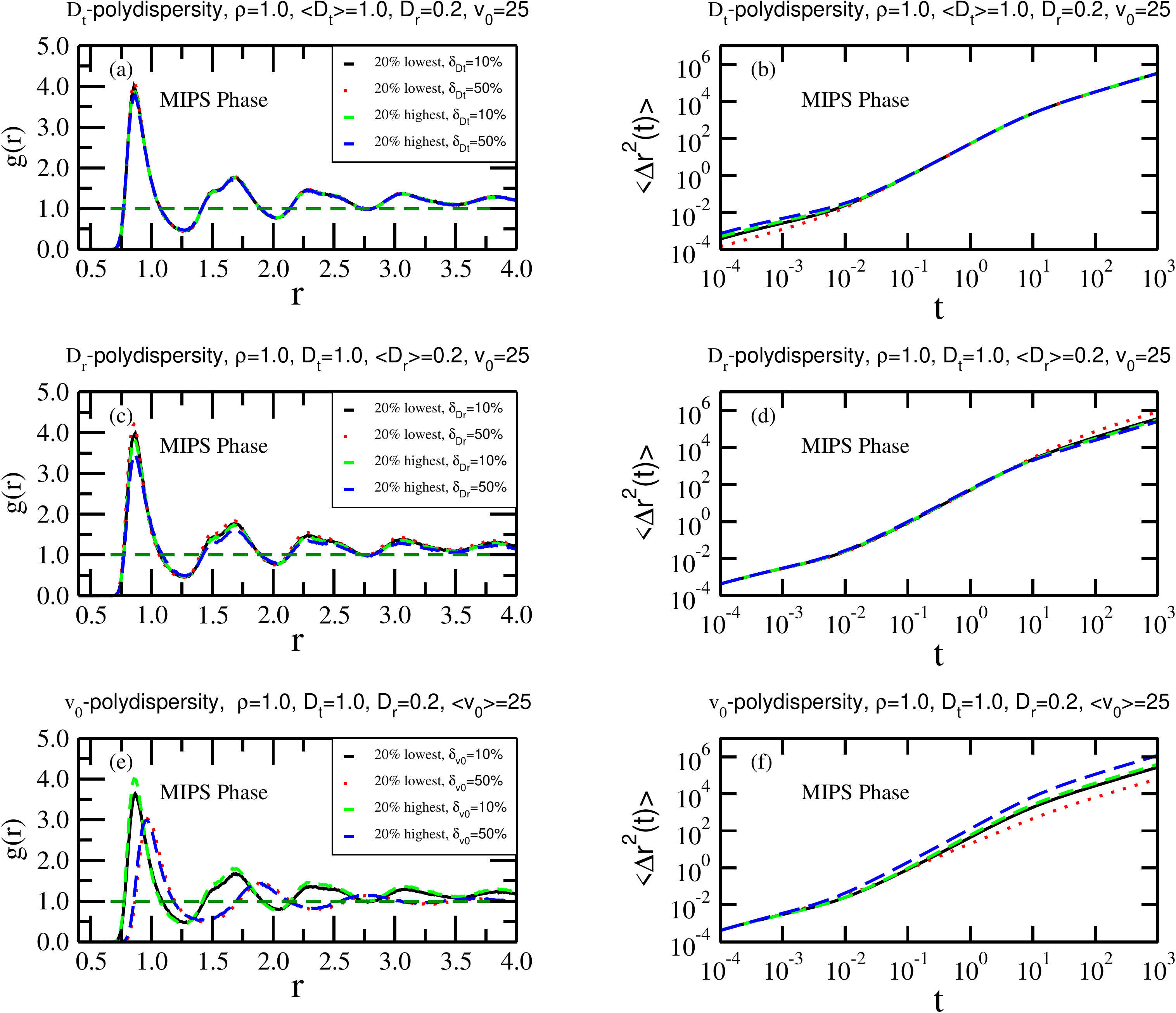}\vspace{2mm}
			\caption{Role of the 20\% lowest (black and red) and 20\% highest (green and blue) active-parameter particles in the MIPS phase of continuously polydisperse systems at 10\% and 50\% polydispersity. 
				(a), (c), and (e) show the RDFs for polydispersity in the $D_t$, $D_r$, and $v_0$ parameters, respectively. 
				(b), (d), and (f) show the corresponding results for the MSD. 
				For the RDFs there is for $D_t$ polydispersity little difference between the lowest and highest active-parameter particles except at the first peak; $D_r$ polydispersity shows a larger but still modest difference, which is most pronounced at 50\% polydispersity. The case of $v_0$ polydispersity shows significant differences between 10\% and 50\% polydispersity, but for each of these values there is only modest variation between the lowest and highest active-parameter particles' RDF.        
				For the MSD the situation is similar to that observed in the homogeneous phase (\fig{fig2}): variation is observed in the short-time data for $D_t$ polydispersity, in the long-time data for $D_r$ polydispersity, and at intermediate and long times for $v_0$ polydispersity.}
			\label{fig4}
		\end{center}
	\end{figure}

	To throw more light on these findings, following the procedure of the homogeneous-phase investigation we identify in \fig{fig4} the contributions to structure and dynamics from the lowest (black and red) and highest (green and blue) parameter particles. Compared to the homogeneous case, there is more variation for all three RDFs, in particular for $D_r$ and $v_0$ polydispersity. In the $D_r$ case, the black and green curves (10\% polydispersity) are close and move in the same direction when increasing to 50\% polydispersity. Interestingly, the average of black and green, as well as of red and blue, is an almost unchanged RDF (\fig{fig3}(c)). The $v_0$ polydispersity case is different: here the 10\% polydispersity curves are similar (black and green), but quite different from the 50\% polydispersity curves (red and blue). This is consistent with the finding of \fig{fig3}(e) and means that the actual value of $v_0$ matters little for the structure surrounding a given particle. We believe this is an effect of the strong interparticle interactions within the MIPS phase that average out the effect of the individually varying $v_0$. At the same time, increasing the degree of $v_0$ polydispersity leads to a considerable broadening of the width of the first peak. Because there is little difference between the low and high $v_0$ RDFs, the picture is very similar to the overall average picture. In fact, at 50\% $v_0$ polydispersity we find that the system becomes almost homogeneous, which is consistent with the findings of Ref. \onlinecite{cas21}. -- In regard to the MSD, the MIPS phase low- and high-parameter findings are similar to those of the homogeneous phase (\fig{fig2}).

	\section{Role of the average potential energy}
	
	\begin{figure}[H]
		\begin{center}
			\includegraphics[width=0.98\textwidth, angle=0]{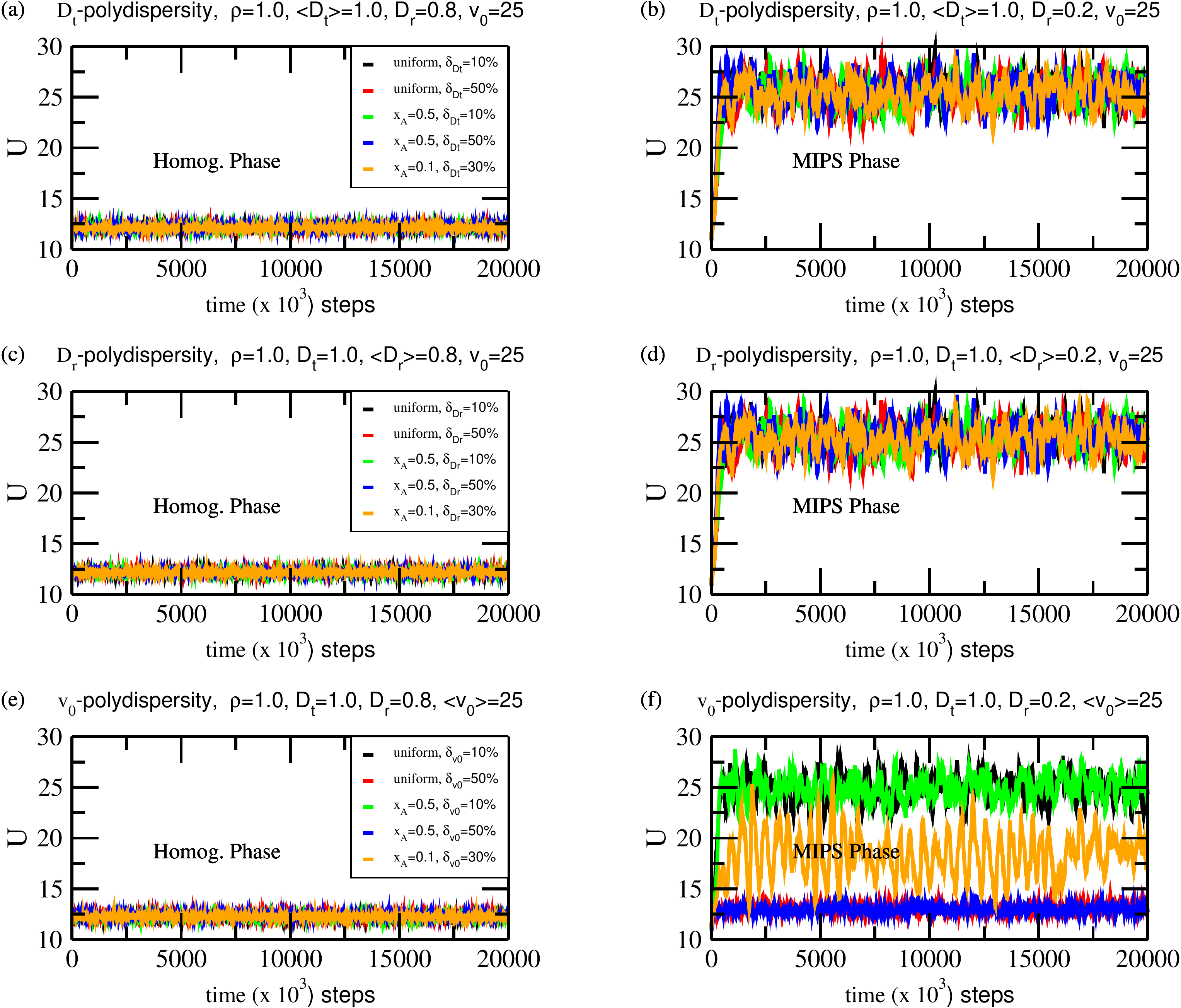}\vspace{2mm}
			\caption{Average potential energy as a function of time during a steady-state simulations. 
				(a), (c), and (e) show data for the homogeneous phase for systems of 10\%, 30\%, and 50\% polydispersity in the $D_t$, $D_r$, and $v_0$ parameters, respectively. 
				(b), (d), and (f) show the corresponding results for the MIPS-phase simulations. 
				Except for the MIPS-phase $v_0$-polydispersity case, the average potential energy is virtually unaffected by the introduction of polydispersity. }
			\label{fig5}
		\end{center}
	\end{figure}
	
	To further illuminate the effect of parameter polydispersity we evaluated the potential energy as a function of time during the simulations (\fig{fig5}). In the homogeneous phase ((a), (c), and (e)) polydispersity has little effect on the average potential energy. This is consistent with the finding that in this phase structure and dynamics are virtually unaffected by the degree of polydispersity (\fig{fig1}). The same applies for the MIPS phase in the $D_t$ and $D_r$ polydispersity cases. Only in the $v_0$-polydispersity MIPS case is the structure significantly affected (\fig{fig3}(e)), which is consistent with the finding that only in this case the average potential energy changes significantly with the degree of polydispersity (\fig{fig5}(f)). At increasing $v_0$ polydispersity the MIPS-phase average potential energy approaches that of the homogeneous phase, which means that the average particle distance increases (the density decreases) with increasing $v_0$ polydispersity. Indeed, in this case the position of the first peak of the RDF was found to increase toward unity (\fig{fig3}(e)), indicating that the MIPS phase gradually fills out the sample area and, eventually, disappears. 
	
	In summary, changes in the average potential energy upon introduction of parameter polydispersity correlate with changes of structure and dynamics. This means that the average potential energy is a convenient ``thermometer'' of changes to the physics.

	\section{Discussion}
	
	It is well known that introducing size polydispersity into active-matter models by varying the characteristic length of the pair potential has a significant effect on both structure and dynamics \cite{ni15,hen20,kum21,sza21,jes23}, just as for passive systems \cite{fre86,sol02,ing15}. This paper investigated the effects of introducing particle-to-particle variations of other parameters of the 2d ABP model with Yukawa pair interactions. With the exception of $v_0$ polydispersity in the MIPS phase, we find surprisingly small effects on the structure and dynamics when polydispersity is introduced such that the average of the parameter in question is kept constant. The cause of this insensitivity to parameter polydispersity is not obvious, but it means that a polydisperse active system in many respects behaves like the homogeneous system of particles with average model parameters, i.e., that a mean-field description applies to a good approximation. While it is easy to understand the significant effects of size polydispersity \cite{ni15,hen20,kum21,sza21,jes23}, we have no physical explanation for the absence of any role of polydispersity in the translational and rotational noise terms, as well as the swim velocity parameter in the homogeneous phase. It often happens in physics that a mean-field description works better than can be justified by simple arguments, and we conclude that this is indeed the case also for parameter polydispersity in active matter. We note that a recent study of different Lennard-Jones (passive) systems showed a similar insensitivity to the introduction of energy polydispersity \cite{ing23}, a result that is also not well understood.
	
	Investigations of other active-matter models should be carried out to determine the generality of our findings. If they are general, the introduction of polydispersity may have applications to instances of non-polydisperse active-matter models for which the system in question is difficult to equilibrate because of extremely long relaxation times \cite{man20,jan22}. The idea is to employ ``activity-induced annealing'' \cite{sha23} for a polydisperse system. As is well known, passive glass-forming polydisperse liquids may be equilibrated by the SWAP algorithm \cite{nin17}. Even though detailed balance does not apply for active matter, SWAP may possibly be applied also for equilibrating an active, single-component highly viscous system \cite{pao22} by proceeding as follows. First, introduce polydispersity into one of the active-model parameters. Then, carry out random particle swaps which according to the above findings will not significantly affect the average structure and dynamics of the system. Finally, remove the artificial polydispersity. Inspired by Ref. \onlinecite{nin17} we conjecture that this will equilibrate the system more quickly than a lengthy simulation.

	\begin{acknowledgments}
		This work was supported by the VILLUM Foundation's \textit{Matter} grant (VIL16515).
	\end{acknowledgments}

\end{document}